\documentclass[
    11pt,
    letterpaper,
    reprint,
    notitlepage,
    superscriptaddress,
    aps,prd
]{revtex4-2}

\UseRawInputEncoding

\usepackage{amsmath,amssymb,amsfonts} 
\usepackage[mathcal]{euscript}
\usepackage{mathtools}

\usepackage{physics}

\usepackage{microtype}
\usepackage{soul}

\usepackage{subdepth}   

\usepackage{graphicx}
\usepackage[svgnames,dvipsnames]{xcolor}
\definecolor{NewBlue}{rgb}{0.1, 0.1, 0.7}
\usepackage[colorlinks,
    linkcolor=FireBrick,
    citecolor=MediumBlue,
    urlcolor=DarkGreen]{hyperref}

\usepackage[capitalise]{cleveref}

\def\nn{\nonumber}

\def\l{\left}
\def\r{\right}
\def\d{{\rm d}}
\def\e{{\rm e}}
\def\k{{\bm k}}
\def\f{\frac}

\def\x{{\bm x}}
\def\p{{\bm p}}
\def\ud{\scriptscriptstyle {\rm UD}}
\def\cm{\rm cm}
\def\rw{{\rm r}}

\newcommand{\ie}{\textit{i.e.}}

\renewcommand{\phi}{\varphi}

\begin{document}

\title{Thermality and athermality in the Unruh effect}

\author{D. Jaffino Stargen}
\email{jaffino@gat.ac.in; jaffinostargend@gmail.com}
\affiliation{Department of Physics, Global Academy of Technology,
Aditya Layout, RR Nagar, Bengaluru, Karnataka 560098, India}
\affiliation{Department of Mechanical Engineering, 
    Massachusetts Institute of Technology, Cambridge, MA 02139, USA}

\author{Vivishek Sudhir}
\affiliation{Department of Mechanical Engineering, 
    Massachusetts Institute of Technology, Cambridge, MA 02139, USA} 
\affiliation{LIGO Laboratory, Massachusetts Institute of Technology, 
    Cambridge, MA 02139, USA}


\begin{abstract}
In idealized treatments of the Unruh effect, a two-level atom is accelerated in a prescribed
classical trajectory through the vacuum of a quantum field --- the Unruh bath --- which causes the 
atom's internal state to thermalize to a temperature proportional to the acceleration.
This happens via emission and absorption of quanta by the atom, leading to a detailed balance between
fluctuations in the Unruh bath and associated dissipation of the atom's internal energy.
In any physical manifestation of the Unruh effect, the center-of-mass (CoM) degree of freedom of the atom
is dynamical, and is therefore coupled to the Unruh bath via momentum recoil during the absorption/emission process. 
We study this scenario and show how the fluctuation-dissipation theorem fails for the CoM degree of freedom, 
while still being upheld for the internal energy. That is,
the CoM of an accelerated atom is not in thermal equilibrium with the Unruh bath, while its internal level can be.
\end{abstract}

\maketitle


\section{Introduction}

The Unruh effect is the prediction that an observer, uniformly accelerated
with respect to an inertial frame, perceives the Minkowski vacuum state of a quantum field to be a thermal bath
at an effective temperature (in natural units) $T_U = a/2\pi$, where 
$a = [-a^{\mu}a_{\mu}]^{1/2}$ is the proper acceleration \cite{Fulling1973,Davies1975,Unruh1976,DeWitt1979,Wald1984,Takagi1986,Crispino2008}.
This effective thermal bath is called the Unruh bath.
Concretely, the observer can be modeled as an idealized two-level atom,
called Unruh-DeWitt (UD) detector, that is uniformly accelerating with respect to
inertial frames, and is linearly coupled to a scalar quantum field \cite{Unruh1976,DeWitt1979,Wald1984,Takagi1986}.
It can be shown that in the frame co-moving with the atom, the interaction between the atom and the field
causes the internal energy of the atom to thermalize with the Unruh bath.
That is, the atom's internal state, over a sufficient period of time, converges to a thermal state 
at temperature, $T_U$, irrespective of its initial state \cite{Merkli2006,Garay2021}.

In idealized models of the Unruh effect it is customary \cite{Unruh1976,DeWitt1979,Wald1984,Takagi1986,Crispino2008,Scully2003,
Mann2011,Scully18,Lochan2020,Sudhir2022,Stargen2022} to neglect the center-of-mass (CoM) motional degree of freedom of the 
accelerated atom. Effectively, these treatments assume the mass of the atom to be infinitely large such that the motional
degree of freedom is effectively decoupled from the dynamics.
A handful of prior work \cite{Casadio1995,Parentani1995,Reznik1998,Casadio1999,Sudhir2021,Kempf2022,Magdalena2022,Magdalena2023}
have investigated the effect of the motional degree of freedom on the Unruh effect. However, the central question of whether
the motional degree of freedom is thermalized to the Unruh bath remains open.
Further, it remains unclear if and how the presence of the motional degree of freedom affects the thermal character of
the accelerated atom's internal two-level system or of the emitted radiation.
Thermality of the radiation emitted by an accelerated atom is a conceptually key requirement that maintains
consistency in black hole thermodynamics \cite{Fulling87,Scully18,WilsGerr24,DanWald25}, 
equivalence principle near horizons \cite{Cand80,SingWil11}, and of quantum field theory
in curved spacetime \cite{Israel76,Sew82}. Thus, the question of thermality of the emitted radiation of a 
\emph{physical (i.e. finitely massive)} accelerated atom needs to be settled.

The purpose of this work is to address this question. We find that the 
center-of-mass (CoM) degree of freedom fails to get thermalized with the Unruh bath;
the effect of the CoM is simply to add mass dependent corrections to the (non-relativistic)
energy gap of the atom's two-level system.
However, and more importantly, the failure of the CoM variables to get thermalized with the Unruh bath
does not affect the thermalization of the internal energy levels.
The rest of this paper proves these assertions in a simple model of a finitely-massive accelerated atom, by attempting to derive
a fluctuation-dissipation type relation \cite{Kubo1957,Haag1967} for the various degrees of freedom of the atom. 

The rest of the paper is organized as follows: 
in \cref{Sec.-II} we state our setup of a finitely massive
non-relativistic UD detector coupled to a massless scalar field, and we show that the Kubo-Martin-Schwinger (KMS) condition \cite{Kubo1957,Schwinger1959,Haag1967}
-- which characterizes systems that are in thermal equilibrium 
-- in the comoving Rindler frame of the detector is not satisfied for all
degrees of freedom of the detector.
In \cref{Sec.-III} we derive the results of \cref{Sec.-II} from an inertial observer's perspective.
In both cases, we elucidate how the fluctuation-dissipation theorem emerges 
in the limit of an infinitely massive detector and derive the leading order 
corrections to it.
In \cref{Sec.-IV}, we show how the fluctuation-dissipation theorem holds
for the internal degrees of freedom of the detector.

\section{Uniformly accelerating atom: co-moving perspective}\label{Sec.-II}

Consider an atom --- modeling the UD detector --- whose ``internal'' two-level system is described by the ground and excited
states $\ket{g}$ and $\ket{e}$ respectively. Let
${\hat H}_{\cal E}=\omega_{0} \sigma_{z}/2$ be the Hamiltonian of these internal states; 
here \mbox{$\sigma_{z} \equiv \l(\dyad{e}{e} - \dyad{g}{g}\r)$}, and $\omega_{0}$ is the energy gap.
The atom interacts with a massless scalar field, ${\hat \phi}$, through an interaction Hamiltonian,
\mbox{${\hat H}_{\rm int} = - \lambda {\hat \mu}(\eta) {\hat \phi}(\eta,x_{\perp},\xi)$},
where ${\hat \mu}$ is the monopole moment of the atom, and $\lambda \ll 1$ is a small coupling constant. 
Here $\xi$ is the coordinate in the comoving (Rindler) frame
along which the atom is uniformly accelerating with a proper four acceleration $a$ (with respect to inertial frames);
$x_\perp$ are the coordinates transverse to $\xi$, and $\eta$ is the coordinate time in the comoving frame.

Apart from the internal energy levels, the atom is free to move along the
transverse coordinates, $x_{\perp}$, while it follows a prescribed trajectory
in the $\xi$-coordinate. Therefore, the Hamiltonian corresponding to its
CoM is \mbox{${\hat H}_{\cm} = {\hat \p}_{\perp}^2/2m$},
where ${\hat \p}_{\perp}=({\hat p}_{x},{\hat p}_{y})$ is the momentum of the detector that is
transverse to the direction of its
acceleration, and they are also canonically conjugate to the position operators,
${\hat x}$ and ${\hat y}$; and $m$ is the mass of the atom. Therefore, the total Hamiltonian of the detector-field system is 
\begin{equation}\label{eqn:H_total}
\begin{split}
    {\hat H}_{\rm total} &= {\hat H}_{\ud} + {\hat H}_{\phi} + {\hat H}_{\rm int} \\
    \text{where,}\qquad 
    {\hat H}_{\ud} &= \f{{\hat p}_{x}^2}{2m}+\f{{\hat p}_{y}^2}{2m} + \f{\omega_{0}}{2} {\hat \sigma}_{z},\\
    {\hat H}_{\phi} &= \int_{0}^{\infty} \d \Omega \int \d \k^2_{\perp}~\Omega~
        {\hat a}^{(\rw)\dagger}_{\Omega,\k_{\perp}} {\hat a}^{(\rw)}_{\Omega,\k_{\perp}},\\
    {\hat H}_{\rm int} &= - \lambda {\hat \mu}(\eta) {\hat \phi}(\eta,{\hat \x}_{\perp},z),
\end{split}
\end{equation}
with $\k_{\perp}=(k_{x},k_{y})$, and the scalar field, ${\hat \phi}$, in the comoving frame
of the atom is expanded in terms of the field modes corresponding to the right Rindler wedge as
\begin{align}
 & \hat{\phi}(x) = \f{1}{2\pi^2} \int_{0}^{\infty} \d \Omega \l(\f{\sinh(\pi\Omega/a)}{a}\r)^{1/2} \nn \\
 &\times \int \d^2 \k_{\perp} K_{i\Omega/a}\l(k_{\perp}/a\r) \nn \\
 &\times 
 \biggl({\hat a}^{(\rw)}_{\Omega,\k_{\perp}} \e^{-i\Omega \eta} \e^{ik_{x}x} \e^{ik_{y}y}
 + {\hat a}^{(\rw)\dagger}_{\Omega,\k_{\perp}} \e^{i\Omega \eta} \e^{-ik_{x}x} \e^{-ik_{y}y}\biggr),
\end{align}
where \mbox{$k_{\perp} \equiv \sqrt{k_{x}^2+k_{y}^2}$}, the ladder operators,
${\hat a}^{(\rw)}_{\Omega,\k_{\perp}},{\hat a}^{(\rw)\dagger}_{\Omega,\k_{\perp}}$, are associated
to the right Rindler wedge, satisfying the commutation relations:
\begin{align}
 [{\hat a}^{(\rw)}_{\Omega,\k_{\perp}},{\hat a}^{(\rw)\dagger}_{\Omega',\k'_{\perp}}]
=\delta(\Omega-\Omega') \delta(\k_{\perp}-\k'_{\perp}),
\end{align}
\begin{align}
 [{\hat a}^{(\rw)}_{\Omega,\k_{\perp}},{\hat a}^{(\rw)}_{\Omega',\k'_{\perp}}]
 = [{\hat a}^{(\rw)\dagger}_{\Omega,\k_{\perp}},{\hat a}^{(\rw)\dagger}_{\Omega',\k'_{\perp}}] = 0,
\end{align}
and the vacuum state corresponding to the field, ${\hat \phi}$, in accelerating frames,
called Rindler vacuum, $\ket{0_{R}}$, is defined as
\mbox{${\hat a}^{(\rw)}_{\Omega,\k_{\perp}} \ket{0_{R}} = 0$}.
\subsection{Linear response regime}
As evident from the structure of the detector-field Hamiltonian in Eq.~(\ref{eqn:H_total}), while the internal
energy levels of the atom are coupled explicitly to the scalar field, ${\hat \phi}$, through its monopole
moment, ${\hat \mu}$,
the CoM variables, $({\hat \x}_{\perp},{\hat \p}_{\perp})$, get coupled to the
field, ${\hat \phi}(\eta,x)$, and the internal energy levels, $\ket{g}$ and $\ket{e}$,
of the atom in an indirect and nontrivial manner, specifically
through the spatial dependence of the field, ${\hat \phi}(\eta,{\hat \x}_{\perp})$.
Therefore, the Unruh bath could thermalize
with the detector variables, \ie, with the variables corresponding to its internal energy levels
and CoM variables, and we clarify this in this section.

A general observable, say, ${\hat X}_{\ud}$, of the UD detector in its comoving Rindler frame evolves
dynamically with respect to the coordinate time,
$\eta$, due to its interaction with the scalar field, and one can find this time evolution,
${\hat X}_{\ud}(\eta)$, by solving the Heisenberg equation of motion
$\dot{\hat X}_{\ud}(\eta) = i [{\hat H}_{\rm total}(\eta),{\hat X}_{\ud}(\eta)]$.
Our interest is in the fluctuations of these observables in the weak-coupling regime, so 
a perturbative treatment, to leading order in the coupling parameter $\lambda$, suffices.
We adopt the formalism of Dalibard, Dupont-Roc, and Cohen-Tannoudji (DDC) \cite{DDC1984}
to elucidate the linear response ${\hat X}_{\ud}(\eta) = {\hat X}_{\ud}^{(0)}(\eta)+\lambda \delta{\hat X}_{\ud}(\eta)+\mathcal{O}(\lambda^2)$, where the linear response part is given by
(see \cref{Appendix:DDC_Formalism} for details),
\begin{align}\label{eqn:LinearResponse-Rindler}
 \delta {\hat X}_{\ud}(\eta) &= -\frac{i\lambda}{2\pi^2} \int_{\eta_{0}}^{\eta} \dd{{\tilde \eta}}
 \int_{0}^{\infty} \d \Omega \l(\f{\sinh(\pi\Omega/a)}{a}\r)^{1/2} \nn \\
 &\times \int \d^2 \k_{\perp} K_{i\Omega/a}(k_{\perp}/a) \nn \\
 &\times \biggl({\hat a}^{(\rw)^{(0)}}_{\Omega,\k_{\perp}}(\eta_{0}) \e^{-i\Omega {\tilde \eta}}
 + {\hat a}^{(\rw)^{(0)}}_{\Omega,-\k_{\perp}}(\eta_{0}) \e^{i\Omega {\tilde \eta}}\biggr) \nn \\
 &\times \l[{\hat \mu}^{(0)}({\tilde \eta}) \e^{ik_{x} {\hat x}^{(0)}({\tilde \eta})}
 \e^{ik_{y} {\hat y}^{(0)}({\tilde \eta})}, {\hat X}_{\ud}^{(0)}(\eta)\r],
\end{align}
where the operators with superscript $"(0)"$, say, ${\hat A}^{(0)}(\eta)$ denotes the time evolution of a given
operator when the interaction between the UD atom and the field is switched off, \ie, $\lambda=0$.
In other words, ${\hat A}^{(0)}(\eta)$ is the solution of the (free) Heisenberg equation of motion,
\mbox{$ \dot{\hat A}^{(0)} = i[{\hat H}_{\ud}(\eta),{\hat A}^{(0)}(\eta)]$}.

In order to study the nature of the fluctuations in these observables, we focus attention 
on their two-point correlation function:
\begin{equation}
    \langle \delta {\hat X}_{\ud}(\eta) \delta {\hat X}_{\ud}(\eta') \rangle
    \equiv \Tr [\delta {\hat X}_{\ud}(\eta) \delta {\hat X}_{\ud}(\eta') \rho^{(0)}],
\end{equation}
where $\rho^{(0)}(\eta_{0})$ is the joint atom-field state at the initial time
$\eta_{0}$. In the following we suppose that
$\rho^{(0)}(\eta) \equiv \rho_{\cal E}^{(0)}(\eta) \otimes \rho_{\cm}^{(0)}(\eta) \otimes \ket{0_{M}}\bra{0_{M}}$,
with $\rho_{\cal E}^{(0)}$ and $\rho_{\cm}^{(0)}$ representing respectively the quantum states corresponding to the internal
energy levels and CoM variables of the detector; and $\ket{0_{M}}$ denotes the Minkowski vacuum of the
scalar field ${\hat \phi}$ in the inertial frame. 

%
%
In the limit $\eta_{0} \to -\infty$, and under the restriction that
$\p_{\perp}^2/2m \ll a$, where $\p_{\perp}$ is the momentum of the detector
transverse to the direction of its acceleration, we evaluate the correlation function
corresponding to the operator, ${\hat X}_{\ud}$, as
\begin{widetext}
\begin{align}\label{Eqn-2}
  & S^{\ud}_{XX}(\eta-\eta') \equiv
  \langle \delta {\hat X}_{\ud}(\eta) \delta {\hat X}_{\ud}(\eta') \rangle
  = \f{\lambda^2 T}{(2\pi)^3} \int_{-\infty}^{\infty} \dd {\tilde u}
 \Theta({\tilde u}) \int_{-\infty}^{\infty} \d \Omega
 \f{(\pi\Omega/a)}{\sinh(\pi\Omega/a)} \e^{\pi\Omega/a} \nn \\
 &\times \Tr_{\rho_{\ud}^{(0)}}
 \biggl\{\biggl[{\hat \sigma}_{-} {\hat \sigma}_{+} \e^{-i\l(\Omega-\omega_{0}+{\hat H}^{(0)}_{\cm}\r) (\eta-\eta')}
 \e^{i\l(\Omega-\omega_{0}+{\hat H}^{(0)}_{\cm}\r) {\tilde u}/a}
 + {\hat \sigma}_{+} {\hat \sigma}_{-} \e^{-i\l(\Omega+\omega_{0}+{\hat H}^{(0)}_{\cm}\r) (\eta-\eta')}
 \e^{i\l(\Omega+\omega_{0}+{\hat H}^{(0)}_{\cm}\r) {\tilde u}/a}\biggr] \nn \\
 &\times {\hat X}_{\ud}^{(0)}(\eta-\eta') {\hat X}_{\ud}^{(0)}(0) \nn \\
 &+ \biggl[{\hat \sigma}_{-} {\hat \sigma}_{+} \e^{-i\l(\Omega+\omega_{0}-{\hat H}^{(0)}_{\cm}\r) (\eta-\eta')}
 \e^{i\l(\Omega+\omega_{0}-{\hat H}^{(0)}_{\cm}\r) {\tilde u}/a}
 + {\hat \sigma}_{+} {\hat \sigma}_{-} \e^{-i\l(\Omega-\omega_{0}-{\hat H}^{(0)}_{\cm}\r) (\eta-\eta')}
 \e^{i\l(\Omega-\omega_{0}-{\hat H}^{(0)}_{\cm}\r) {\tilde u}/a}\biggr] \nn \\
 &\times {\hat X}_{\ud}^{(0)}(\eta'-\eta) {\hat X}_{\ud}^{(0)}(0)\biggr\},
\end{align}
\end{widetext}
where $T$ is the total time of interaction of the UD detector with the quantum field,
and ${\hat \sigma}_{+}=\ket{e}\bra{g}$ and ${\hat \sigma}_{-}=\ket{g}\bra{e}$.

Note that, as evident from Eq.(\ref{Eqn-2}), the quantum fluctuations of the scalar field in
the comoving Rindler frame of
the atom induces a stationary quantum process on the observables, ${\hat X}_{\ud}$, of the detector
(see Appendix-\ref{Appendix:SXX} for more details).
It needs to be ensured whether the CoM variables and the internal energy levels of the detector
establish a thermal equilibrium with the Unruh bath,
and we address this in the rest of this section.
\subsection{KMS condition: Do CoM variables get thermalized with the Unruh bath?}
To ensure whether the detector is in thermal equilibrium with the Unruh bath,
it is sufficient to show that the correlation function, $S^{\ud}(\eta-\eta')$,
corresponding to the detector observables,
${\hat X}_{\ud}$, satisfy KMS condition, \ie,
\begin{align}
 S_{XX}^{\ud}(\eta-\eta') = S^{\ud}_{XX}(\eta'-\eta-2\pi i/a).
\end{align}
From the expression for correlation function, $S^{\ud}(\eta-\eta')$, in Eq.~(\ref{Eqn-2}), we obtain
\begin{widetext}
\begin{align}\label{Eqn-3}
 & S^{\ud}_{XX}(\eta-\eta') - S^{\ud}_{XX}(\eta'-\eta-2\pi i/a) = \f{\lambda^2 T}{(2\pi)^3}
  \int_{-\infty}^{\infty} \dd {\tilde u}
 \Theta({\tilde u}) \int_{-\infty}^{\infty} \d \Omega (\pi\Omega/a) \coth(\pi\Omega/a) \nn \\
 &\times \Tr_{\rho_{\ud}^{(0)}}
 \biggl\{\biggl[\sigma_{-} \sigma_{+} \e^{-i\l(\Omega-\omega_{0}+{\hat H}^{(0)}_{\cm}\r) (\eta-\eta')}
 \e^{i\l(\Omega-\omega_{0}+{\hat H}^{(0)}_{\cm}\r) {\tilde u}/a}
 + \sigma_{+} \sigma_{-} \e^{-i\l(\Omega+\omega_{0}+{\hat H}^{(0)}_{\cm}\r) (\eta-\eta')}
 \e^{i\l(\Omega+\omega_{0}+{\hat H}^{(0)}_{\cm}\r) {\tilde u}/a}\biggr] \nn \\
 &\times \biggl({\hat X}_{\ud}^{(0)}(\eta-\eta') {\hat X}_{\ud}^{(0)}(0)
 - \e^{-2\pi{\hat H}^{(0)}_{\cm}/a}
 {\hat X}_{\ud}^{(0)}(\eta-\eta') \e^{2\pi{\hat H}^{(0)}_{\cm}/a} \times {\hat X}_{\ud}^{(0)}(0)\biggr) \nn \\
 &+ \biggl[\sigma_{-} \sigma_{+} \e^{-i\l(\Omega+\omega_{0}-{\hat H}^{(0)}_{\cm}\r) (\eta-\eta')}
 \e^{i\l(\Omega+\omega_{0}-{\hat H}^{(0)}_{\cm}\r) {\tilde u}/a}
 + \sigma_{+} \sigma_{-} \e^{-i\l(\Omega-\omega_{0}-{\hat H}^{(0)}_{\cm}\r) (\eta-\eta')}
 \e^{i\l(\Omega-\omega_{0}-{\hat H}^{(0)}_{\cm}\r) {\tilde u}/a}\biggr] \nn \\
 &\times \biggl({\hat X}_{\ud}^{(0)}(\eta'-\eta) {\hat X}_{\ud}^{(0)}(0)
 - \e^{2\pi{\hat H}^{(0)}_{\cm}/a} {\hat X}_{\ud}^{(0)}(\eta'-\eta) \e^{-2\pi{\hat H}^{(0)}_{\cm}/a}
 \times {\hat X}_{\ud}^{(0)}(0)\biggr)\biggr\}.
\end{align}

\end{widetext}

Note that the appearance of the CoM Hamiltonian, ${\hat H}_{\cm}$, along with the energy gap, $\omega_{0}$
of the detector, \ie, as $\omega_{0}\pm{\hat H}_{\cm}$ in the expression for correlation function,
$S^{\ud}_{XX}(\eta-\eta')$, in Eq.~(\ref{Eqn-3}) indicates that the quantum state, $\rho_{\ud}$,
corresponding to the
detector cannot establish thermal equilibrium with the Unruh bath, if the detector's CoM is taken
into consideration.
However, if one restricts oneself to a class of operators corresponding to the atom, say,
${\hat X'}_{\ud}$, that commutes with the CoM Hamiltonian, ${\hat H}_{\cm}$, of the detector \ie,
$[{\hat X'}_{\ud},{\hat H}_{\cm}]=0$, then it is straightforward to deduce from Eq.~(\ref{Eqn-3}) that
\begin{align}\label{eqn:KMS-Rindler}
 S^{\ud}_{X'X'}(\eta-\eta') = S^{\ud}_{X'X'}(\eta'-\eta-2\pi i/a).
\end{align}

Making use of the fact that the UD detector gets thermalized with the Unruh bath for a restricted set of observables
that satisfy the operator expression, $[{\hat X'}_{\ud},{\hat H}_{\cm}]=0$,
we establish the fluctuation-dissipation relation corresponding to the UD operators,
${\hat X}'_{\ud}$, in Sec.~(\ref{Sec.-IV}).
\section{Uniformly accelerating atom: Inertial observer's perspective}\label{Sec.-III}
With respect to an inertial frame, the interaction Hamiltonian, ${\hat H}^{({\rm I})}_{\rm int}$, corresponding
to the interaction between the
UD detector and the massless scalar field, ${\hat \phi}$, takes the form
\mbox{${\hat H}^{({\rm I})}_{\rm int} = - \lambda {\hat \mu} {\hat \phi}(t,\x_{\perp},z)$},
where $t$ is the coordinate
time with respect to the inertial frame, and $\x_{\perp}=(x,y)$ are the coordinates that are
transverse to the $z$-coordinate. If the detector is assumed to be uniformly accelerating along the
$z$-axis,
then the corresponding trajectory information can be stated as
\begin{align}
 t=a^{-1} \sinh a\tau, ~~ z=a^{-1} \cosh a\tau,
\end{align}
where $a$ and $\tau$ denote the proper acceleration and proper time of the detector, respectively.

The detector is assumed to be moving along $z$-axis in a given trajectory, but
is free to move in directions transverse to $z$-axis. Therefore, the transverse coordinates,
$\x_{\perp}$, remain
unaffected under coordinate transformations between the inertial and Rindler frames. Therefore,
apart from the internal energy levels, $\ket{g}$ and $\ket{e}$, the detector possesses the degrees of freedom
corresponding to its CoM too, \ie,
$x$ and $y$, and the corresponding canonically conjugate momentum variables are respectively $p_{x}$ and $p_{y}$.
This leads to the Hamiltonian corresponding to the UD detector in the inertial frame as
\begin{align}
 {\hat H}^{({\rm I})}_{\ud} = \f{{\hat p}_{x}^2}{2m}+\f{{\hat p}_{y}^2}{2m}
 + \f{\omega_{0}}{2} {\hat \sigma}_{z},
\end{align}
where $m$ and $\omega_{0}$ are respectively the mass and energy gap of the detector, and
${\hat \sigma}_{z} \equiv \ket{e}\bra{e}-\ket{g}\bra{g}$.
Therefore, the total Hamiltonian corresponding to the detector-field system in the inertial frame
can be written as
${\hat H}^{({\rm I})}_{\rm total} = {\hat H}^{({\rm I})}_{\ud} + {\hat H}^{({\rm I})}_{\phi}
+ {\hat H}^{({\rm I})}_{\rm int}$,
where
\begin{align}
 {\hat H}^{({\rm I})}_{\phi} = \int \d^3 \k \omega_{k} {\hat a}^{\dagger}_{\k} {\hat a}_{\k},
\end{align}
and the ladder operators,
\mbox{${\hat a}_{\k},{\hat a}^{\dagger}_{\k}$}, satisfy
\begin{align}
 [{\hat a}_{\k},{\hat a}^{\dagger}_{\k'}]=\delta(\k-\k'),~~
%
 [{\hat a}_{\k},{\hat a}_{\k'}]=[{\hat a}^{\dagger}_{\k},{\hat a}^{\dagger}_{\k'}]=0,
\end{align}
with the Minkowski vacuum, $\ket{0_{M}}$, defined as \mbox{${\hat a}_{\k} \ket{0_{M}} = 0$}.
\subsection{Linear response regime in inertial frame}
As in previous section, a general observable, say, ${\hat Y}_{\ud}$, corresponding to the
UD detector in the inertial
frame evolves dynamically with respect to the proper time,
$\tau$, due to its interaction with the scalar field, and one can find this time evolution,
${\hat Y}_{\ud}(\tau)$, by solving the Heisenberg equation of motion
\begin{align}
 \f{\d {\hat Y}_{\ud}}{\d t} = i\l[{\hat H}^{({\rm I})}_{\rm total},{\hat Y}_{\ud}(\tau)\r].
\end{align}
For the detector-field Hamiltonian, ${\hat H}^{({\rm I})}_{\rm total}$, in the inertial frame,
we employ the DDC formalism \cite{DDC1984} to arrive at the perturbative solution of the form,
${\hat Y}_{\ud}(\tau) = {\hat Y}_{\ud}^{(0)}(\tau)+\delta{\hat Y}_{\ud}(\tau)+\order{\lambda^2}$, with
\begin{align}
 \delta {\hat Y}_{\ud}(\tau) &= -\frac{i\lambda}{(2\pi)^{3/2}} \int_{\tau_{0}}^{\tau} \dd\tau
 \int \f{\d^3 \k}{\sqrt{2\omega_{k}}} \e^{ik_{z} {\tilde z}(\tau)} \nn \\
 &\times \biggl({\hat a}^{(0)}_{\k}(t_{0}) \e^{-i\omega_{k} {\tilde t}(\tau)}
 + {\hat a}^{(0)}_{-\k}(t_{0}) \e^{i\omega_{k} {\tilde t}(\tau)}\biggr) \nn \\
 &\times \l[{\hat \mu}^{(0)}(\tau) \e^{ik_{x} {\hat x}^{(0)}(\tau)}
 \e^{ik_{y} {\hat y}^{(0)}(\tau)}, {\hat Y}_{\ud}^{(0)}(\tau)\r],
\end{align}
where the operators with superscript $"(0)"$ denotes the time evolution of the corresponding
operator when the interaction is switched off, \ie, $\lambda=0$, and $\tau_{0}$ denotes an initial
time at which the UD detector starts interacting with the scalar field, ${\hat \phi}$.

We Suppose, at the initial time, $\tau_{0}$, the detector-field system starts in an initial state,
$\rho_{0}=\rho^{(0)}(\tau_{0})$, where
$\rho^{(0)}(\tau) \equiv \rho_{\cal E}^{(0)}(\tau) \otimes \rho_{\cm}^{(0)}(\tau)
\otimes \ket{0_{M}}\bra{0_{M}}$,
with $\rho_{\cal E}^{(0)}$ and $\rho_{\cm}^{(0)}$ represent, respectively,
the quantum states corresponding to the internal
energy levels and CoM variables of the detector; and $\ket{0_{M}}$ denotes the Minkowski vacuum of the
scalar field, ${\hat \phi}$, in the inertial frame. We calculate the correlation function
corresponding to the operator, ${\hat Y}_{\ud}$, of the detector with respect
to the unperturbed state, $\rho^{0}(t)$,
in the limit $\tau_{0} \to -\infty$, and under the restriction that $\p_{\perp}^2/2m \ll a$,
we find the expression for the correlation function as (see Appendix-\ref{Appendix:SYY} for more details)

\begin{widetext}
\begin{align}\label{eqn:LabCF}
 & S^{\ud}_{YY}(\tau-\tau') \equiv \langle \delta {\hat Y}_{\ud}(\tau) \delta {\hat Y}_{\ud}(\tau') \rangle
 = \f{\lambda^2 T}{(2\pi)^3} \int_{-\infty}^{\infty} \dd {\tilde u}
 \Theta({\tilde u}) \int_{-\infty}^{\infty} \d \Omega
 \f{(\pi\Omega/a)}{\sinh(\pi\Omega/a)} \e^{\pi\Omega/a} \nn \\
 &\times \Tr_{\rho_{\ud}^{(0)}}
 \biggl\{\biggl({\hat \sigma}_{-} {\hat \sigma}_{+} \e^{-i\l(\Omega-\omega_{0}+{\hat H}^{(0)}_{\cm}\r) (\tau-\tau')}
 \e^{i\l(\Omega-\omega_{0}+{\hat H}^{(0)}_{\cm}\r) {\tilde u}/a}
 + {\hat \sigma}_{+} {\hat \sigma}_{-} \e^{-i\l(\Omega+\omega_{0}+{\hat H}^{(0)}_{\cm}\r) (\tau-\tau')}
 \e^{i\l(\Omega+\omega_{0}+{\hat H}^{(0)}_{\cm}\r) {\tilde u}/a}\biggr) \nn \\
 &\times {\hat Y}_{\ud}^{(0)}(\tau-\tau') {\hat Y}_{\ud}^{(0)}(0) \nn \\
 &+ \biggl({\hat \sigma}_{-} {\hat \sigma}_{+} \e^{-i\l(\Omega+\omega_{0}-{\hat H}^{(0)}_{\cm}\r) (\tau-\tau')}
 \e^{i\l(\Omega+\omega_{0}-{\hat H}^{(0)}_{\cm}\r) {\tilde u}/a}
 + {\hat \sigma}_{+} {\hat \sigma}_{-} \e^{-i\l(\Omega-\omega_{0}-{\hat H}^{(0)}_{\cm}\r) (\tau-\tau')}
 \e^{i\l(\Omega-\omega_{0}-{\hat H}^{(0)}_{\cm}\r) {\tilde u}/a}\biggr) \nn \\
 &\times {\hat Y}_{\ud}^{(0)}(\tau'-\tau) {\hat Y}_{\ud}^{(0)}(0)\biggr\}.
\end{align}
\end{widetext}

Note that the structural form of the correlation function, $S^{\ud}_{YY}$, in Laboratory frame
(Eq.(\ref{eqn:LabCF})) matches exactly with the expression for correlation function,
$S^{\ud}_{XX}$, in the comoving Rindler frame of the detector
in Eq.(\ref{Eqn-2}). Therefore, the arguments that
we made for $S^{\ud}_{XX}$ in section.~\ref{Sec.-II} also applies to the correlation function, $S^{\ud}_{YY}$,
in the laboratory frame. Particularly, as for the case of the correlation function,
$S^{\ud}_{XX}$, in the comoving
Rindler frame of the detector, the correlation function, $S^{\ud}_{YY}$, in the laboratory frame also does not
in general satisfy the KMS condition, which means that the quantum state corresponding to the UD detector,
$\rho_{\ud}$, do not necessarily
get thermalized with the Unruh bath in the laboratory perspective too.
However, if one restricts oneself to a class of operators, say, ${\hat Y'}_{\ud}$, corresponding to the
UD detector in the laboratory frame that commutes with the CoM Hamiltonian, ${\hat H}^{(I)}_{\cm}$, \ie,
$[{\hat Y'}_{\ud},{\hat H}^{(I)}_{\cm}]=0$, then it is straightforward to show, as in section.\ref{Sec.-II}, that
\begin{align}\label{eqn:KMS-Lab}
 S^{\ud}_{Y'Y'}(\tau-\tau') = S^{\ud}_{Y'Y'}(\tau'-\tau-2\pi i/a).
\end{align}

Making use of the fact that the UD detector gets thermalized with the Unruh bath for a restricted set
of observables that satisfy the operator expression, $[{\hat Y'}_{\ud},{\hat H}_{\cm}]=0$,
we establish the fluctuation-dissipation relation corresponding to the UD operators,
${\hat Y}'_{\ud}$, in Sec.~(\ref{Sec.-IV}).
\section{Thermalization through internal energy levels of the UD detector}\label{Sec.-IV}
Since the KMS condition corresponding to the UD detector remains valid in both the comoving
Rindler frame, \ie, Eq.(\ref{eqn:KMS-Rindler}), and the laboratory frame, \ie, Eq.(\ref{eqn:KMS-Lab}), of the detector,
particularly for the operators that commute with the CoM Hamiltonian, ${\hat H}_{\cm}$, in this section, we
establish the fluctuation-dissipation theorem in both
the comoving Rindler frame and the laboratory frame of the detector in an unified manner.

Making use of the expressions for
correlation function, $S_{XX}(\eta-\eta')$, in the comoving Rindler frame in Eq.(\ref{Eqn-2}),
and the expression for the correlation function, $S_{YY}(\tau-\tau')$, in the laboratory frame of the detector
in Eq.(\ref{eqn:LabCF}), we obtain
\begin{align}\label{eqn:FDT-1}
 & S^{\ud}_{ZZ}(u)-S^{\ud}_{ZZ}(-u) = \f{\lambda^2}{2\pi} \Theta(u) \nn \\
 &\times \Tr_{\rho_{\ud}^{(0)}} \biggl\{(\omega_{0}-{\hat H}^{(0)}_{\cm})
 \times \sigma_{-}\sigma_{+} \l[{\hat Z}_{\ud}^{(0)}(u),{\hat Z}_{\ud}^{(0)}(0)\r] \nn \\
 &- (\omega_{0}+{\hat H}^{(0)}_{\cm})
 \times \sigma_{+}\sigma_{-} \l[{\hat Z}_{\ud}^{(0)}(u),{\hat Z}_{\ud}^{(0)}(0)\r]\biggr\},
\end{align}
where ${\hat Z}_{\ud}$ represents the operators, ${\hat X}_{\ud}$ (${\hat Y}_{\ud}$),
corresponding to the detector,
and $u=\eta-\eta'$ $(u=\tau-\tau')$ when the expression is written in
the comoving Rindler frame (laboratory frame) of the detector.

Since the KMS condition remains valid only for the operators $X'_{\ud}$ ($Y'_{\ud}$) in the comoving
Rindler frame (laboratory frame) of the detector, as discussed in sections. \ref{Sec.-II} and
\ref{Sec.-III}, one might expect that the expression in Eq.(\ref{eqn:FDT-1}) when restricted to operators
$X'_{\ud}$ and $Y'_{\ud}$ must lead to fluctuation-dissipation relation. But this is not the case,
since the presence of CoM Hamiltonian, ${\hat H}_{\cm}$, in Eq.(\ref{eqn:FDT-1}) obstructs one from identifying
a term in its RHS that denotes the susceptibility of the detector.

Supposing we restrict the expression in Eq.(\ref{eqn:FDT-1}) to the operators,
${\hat Z}_{{\cal E}}$, that correspond only to the internal degrees of freedom of the detector, then
we obtain
\begin{align}
 & S^{{\cal E}}_{ZZ}(u)-S^{{\cal E}}_{ZZ}(-u) = \f{\lambda^2}{2\pi} \Theta(u) \nn \\
 &\times \biggl\{(\omega_{0}-E_{\cm})
 \times \Tr_{\rho_{{\cal E}}^{(0)}}
 \l(\sigma_{-}\sigma_{+} \l[{\hat Z}_{{\cal E}}^{(0)}(u),{\hat Z}_{{\cal E}}^{(0)}(0)\r]\r) \nn \\
 &- (\omega_{0}+E_{\cm}) \times \Tr_{\rho_{{\cal E}}^{(0)}}
 \l(\sigma_{+}\sigma_{-} \l[{\hat Z}_{{\cal E}}^{(0)}(u),{\hat Z}_{{\cal E}}^{(0)}(0)\r]\r)\biggr\},
\end{align}
where $E_{\cm}=\Tr_{\rho^{(0)}_{\cm}}\l(\hat p^2/2m\r)$.
Assuming the detector starts from its ground state, $\ket{g}$, we obtain
\begin{align}\label{eqn:FDT-2}
 S^{{\cal E}}_{ZZ}(u)-S^{{\cal E}}_{ZZ}(-u) &= \f{2\lambda^2}{(2\pi)^{2}}
 \pi (\omega_{0}-E_{\cm}) \nn \\
 &\times \f{1}{2i} \l[\chi(u)-\chi(-u)\r],
\end{align}
where
\begin{align}
 \chi(u)=i\Theta(u) \bra{g} \l[{\hat Z}_{{\cal E}}^{(0)}(u),{\hat Z}_{{\cal E}}^{(0)}(0)\r] \ket{g},
\end{align}
is the susceptibility corresponding to the internal energy levels of the UD detector.
Fourier transforming the expression in Eq.(\ref{eqn:FDT-2}), and also making use of the KMS condition,
$S^{{\cal E}}_{ZZ}(u)=S^{{\cal E}}_{ZZ}(-u-2\pi i/a)$, corresponding to the internal energy levels of
the detector, we obtain
\begin{align}
 S^{{\cal E}}_{ZZ}[\Omega] = \f{\lambda^2}{2\pi}
 \f{(\omega_{0}-E_{\cm})}{\l(1-\e^{-2\pi\Omega/a}\r)} \Im\chi[\Omega],
\end{align}
where the Fourier transform of a function, say $f(t)$, is defined as
$f[\Omega]=\int_{-\infty}^{+\infty} \d t~\e^{-i\Omega t} f(t)$.

It is evident from Eq.(\ref{eqn:FDT-1}) that when the CoM variables are incorporated with the internal energy
levels of the detector,
maintaining the validity of the KMS condition corresponding to the CoM variables is not necessarily enough to ensure
the validity of the corresponding fluctuation-dissipation theorem,
which means that the CoM degrees of freedom cannot thermalize
with the Unruh bath. Therefore, though the absorption and emission rates of the detector are nontrivially modified
by the effects of CoM variables
\cite{Casadio1995,Parentani1995,Reznik1998,Sudhir2021,Magdalena2022,Magdalena2023},
they do not satisfy the detailed balance condition, which in turn leads to the
violation of the KMS condition corresponding to the CoM variables.

Testing Unruh effect means experimentally probing the thermal aspects of the Unruh bath,
particularly through a
quantum probe, called UD detector, where the observables corresponding to the detector are required
to get thermalized with the Unruh bath. In other words,
any detector observable that fail to arrive at a thermal state due to its interaction with the Unruh bath
cannot be employed in an experiment that aims to test Unruh effect. Therefore, since the quantum state
corresponding to the CoM observables of the detector do not arrive at a thermal state due to its interaction
with the Unruh bath, the CoM observables, \ie, position and momentum variables,
$({\hat \x}_{\perp},{\hat \p}_{\perp})$, cannot be employed as suitable observables in any experimental
attempt that tests Unruh effect.

Interestingly, the fact that the quantum state corresponding to the internal energy levels of the
detector arrives at a thermal state with inverse temperature, $\beta=2\pi/a$ \cite{Merkli2006}, is not affected
by the incorporation of CoM variables of the detector. This means that the variables corresponding to the internal
degrees of freedom of the atom are indeed the robust ones that can be employed in experimental setups that aim
to test the thermal nature of the Unruh bath.
\section{Discussions}\label{Sec.-V}
In this work we calculated the correlation functions corresponding to the observables of a uniformly
accelerating UD detector that is coupled to the massless scalar field. We find that the correlation functions
corresponding to the UD detector observables, such as
${\hat \x}_{\perp}$ and ${\hat \p}_{\perp}$,
do not satisfy the KMS condition in general,
\ie, only the correlation functions, $S^{\ud}_{Z'Z'}$, corresponding to the observables, $Z'_{\ud}$, that
commute with the CoM Hamiltonian, ${\hat H}_{\cm}$, of the detector satisfy the KMS condition.
We note that the
validity of the KMS condition does not necessarily ensure the detector observables to satisfy the
fluctuation-dissipation relation, and the fluctuation-dissipation theorem remains
valid only for the operators corresponding to the internal energy levels of the detector,
\ie, $\sigma_{+}$, $\sigma_{-}$, and $\sigma_{z}$.

Testing the existence of Unruh effect amounts to experimentally probing the thermal
aspects of Unruh bath through
a set of relevant observables corresponding to a quantum probe (UD detector), provided the quantum probe
maintains a thermal equilibrium with the Unruh bath. For our case of the uniformly
accelerating finitely massive
UD detector, we find that only the operators corresponding to the internal energy levels of the detector, \ie,
$\sigma_{+}$, $\sigma_{-}$, and $\sigma_{z}$, satisfy the fluctuation-dissipation theorem. Therefore, for
the case of a non-relativistic finitely massive UD detector, the suitable observable
that aim to probe the existence
of the Unruh effect are the observables corresponding to the internal energy levels of the detector,
\ie, $\sigma_{+}$, $\sigma_{-}$, and $\sigma_{z}$.

\bibliography{refs}

\appendix
\begin{widetext}
\section{Calculation details of Eq.(\ref{eqn:LinearResponse-Rindler})}\label{Appendix:DDC_Formalism}
In this appendix we state out the relevant aspects of the DDC formalism \cite{DDC1984} that is required to
arrive at Eq.(\ref{eqn:LinearResponse-Rindler}). Suppose a system, $S$, with Hamiltonian, ${\hat H}_{S}$, is
weakly interacting with an environment, also called reservoir, $R$, with Hamiltonian, ${\hat H}_{R}$, and the
corresponding interaction Hamiltonian is ${\hat H}_{\rm int}=-\lambda\sum_{i} {\hat S}_{i} {\hat R}_{i}$, where
the parameter, $\lambda$, is assumed to be small, \ie, $\lambda\ll1$. Therefore, the total Hamiltonian of the
system can be written as
\begin{equation}
 {\hat H} = {\hat H}_{S} + {\hat H}_{R} - \lambda \sum_{i} {\hat S}_{i} {\hat R}_{i},
\end{equation}
where ${\hat S}_{i}$ and ${\hat R}_{i}$ respectively denote the operators corresponding to the system and reservoir.

Any general operator, say, $O(\tau)$, in this system-reservoir setup follows the Heisenberg equation of motion as
\begin{eqnarray}
 \f{\d {\hat O}(\tau)}{\d \tau} = i [{\hat H}_{S} + {\hat H}_{R},{\hat O}]
 - i\lambda \sum_{i} [{\hat R}_{i}S_{i},{\hat O}],
\end{eqnarray}
and one can calculate the solution of this equation as a power series in $\lambda$. For this purpose,
it is useful to introduce the operators, ${\hat q}_{ab}=\dyad{a}{b}$ and ${\hat Q}_{AB}=\dyad{A}{B}$,
where $\ket{a},\ket{b}$ are eigenstates of ${\hat H}_{S}$ with energy eigenvalues
$\varepsilon_{a}, \varepsilon_{b}$, and $\ket{A},\ket{B}$ are eigenstates of ${\hat H}_{R}$ with
energy eigenvalues $E_{A}, E_{B}$. Therefore, the Heisenberg equation corresponding to ${\hat q}_{AB}$ reads
\begin{equation}
 \derivative{{\hat q}_{ab}(\tau)}{\tau}
 = i \Omega_{ab} {\hat q}_{ab}(\tau) - i\lambda
 \sum_{i} {\hat R}_{i}(\tau) \comm{{\hat S}_{i}(\tau)}{{\hat q}_{ab}(\tau)},
\end{equation}
where $\Omega_{ab} = (\varepsilon_{a}-\varepsilon_{b})$, and integrating the above equation one obtains
\begin{equation}
 {\hat q}_{ab}(\tau) =  {\hat q}_{AB}(\tau_0) e^{i\Omega_{ab}(\tau-\tau_0)} + \delta {\hat q}_{ab}(\tau)
 + \order{\lambda^2},
\end{equation}
where
\begin{equation}
 \delta {\hat q}_{ab}(\tau) = - i\lambda \sum_i \int_{\tau_0}^{\tau} \dd{\tau'}
 e^{-i\Omega_{ab}(\tau'-\tau)} {\hat R}_{i}(\tau')
 \comm{{\hat S}_{i}(\tau')}{{\hat q}_{ab}(\tau')}
 + \order{g^2}.
\end{equation}

Since any general system operator, say, ${\hat X}$, can be written as
\begin{equation}
 {\hat X}(\tau) = \sum_{ab} q_{ab}(\tau) \matrixel{a}{{\hat X}}{b},
\end{equation}
the Heisenberg equation of motion corresponding to the system operator, ${\hat X}$, \ie,
$\d {\hat X}(\tau)/\d \tau = i [{\hat H},{\hat X}]$, can be evaluated to be
\begin{equation}
 {\hat X}(\tau) = {\hat X}^{(0)}(\tau) + \delta {\hat X}(\tau)  + \order{\lambda^2},
\end{equation}
where ${\hat X}^{(0)}(\tau)$ satisfies
$\d {\hat X}^{(0)}(\tau)/\d \tau = i [{\hat H}_{S},{\hat X}^{(0)}]$, and
\begin{align}
 \delta {\hat X}(\tau) = - \frac{ig}{\hbar} \int_{\tau_0}^{\tau} \dd{\tau'}
 \sum_{i} {\hat R}^{(0)}_{i}(\tau')
 \comm{{\hat S}^{(0)}_{i}(\tau')}{{\hat X}^{(0)}(\tau)}.
\end{align}

Therefore, applying the above formalism to our detector-field setup with the total Hamiltonian,
${\hat H}_{\rm total}$, in the comoving Rindler frame of the detector, \ie,
\begin{align}
 {\hat H}_{\rm total} &= \underbrace{\f{{\hat p}_{x}^2}{2m} + \f{{\hat p}_{y}^2}{2m}
 + \f{\omega_{0}}{2} \l(\dyad{e}{e} - \dyad{g}{g}\r)}_{{\hat H}_{S}}
 + \underbrace{\int_{0}^{\infty} \d \Omega \int \d k_{x} \d k_{y}~\Omega
 {\hat a}^{(\rw)\dagger}_{\Omega,\k_{\perp}} {\hat a}^{(\rw)}_{\Omega,\k_{\perp}}}_{{\hat H}_{R}} \nn \\
 &- g \int_{0}^{\infty} \d \Omega \int \d k_{x} \d k_{y}~
 \biggl(\underbrace{{\hat a}^{(\rw)}_{\Omega,\k_{\perp}}}_{R_{i}}
 \times \underbrace{{\hat \mu}(\eta) u^{(\rw)}_{\Omega,\k_{\perp}}(\eta,{\hat x},\hat{y})}_{S_{i}}
 + \underbrace{{\hat a}^{(\rw)\dagger}_{\Omega,\k_{\perp}}}_{R_{i}}
 \times \underbrace{{\hat \mu}(\eta) u^{(\rw)*}_{\Omega,\k_{\perp}}(\eta,{\hat x},\hat{y})}_{S_{i}}\biggr),
\end{align}
we obtain
\begin{align}
 \delta {\hat X}_{\ud}(\eta) &= -\frac{i\lambda}{2\pi^2} \int_{\eta_{0}}^{\eta} \dd{{\tilde \eta}}
 \int_{0}^{\infty} \d \Omega \l(\f{\sinh(\pi\Omega/a)}{a}\r)^{1/2}
 \int \d^2 \k_{\perp} K_{i\Omega/a}(k_{\perp}/a) \nn \\
 &\times \biggl({\hat a}^{(\rw)^{(0)}}_{\Omega,\k_{\perp}}(\eta_{0}) \e^{-i\Omega {\tilde \eta}}
 + {\hat a}^{(\rw)^{(0)}}_{\Omega,-\k_{\perp}}(\eta_{0}) \e^{i\Omega {\tilde \eta}}\biggr)
 \times \l[{\hat \mu}^{(0)}({\tilde \eta}) \e^{ik_{x} {\hat x}^{(0)}({\tilde \eta})}
 \e^{ik_{y} {\hat y}^{(0)}({\tilde \eta})}, {\hat X}_{\ud}^{(0)}(\eta)\r].
\end{align}
\section{Calculation details of Eq.(\ref{Eqn-2})}\label{Appendix:SXX}
Making use of the expression for the operator, ${\hat X}_{\ud}$, in the linear response regime, \ie,
$\delta {\hat X}_{\ud}(\eta)$, in Eq.\ref{eqn:LinearResponse-Rindler}, and making use of the expressions below
\begin{align}
 \langle 0_{M}| {\hat a}^{(\rw)}_{\Omega,\k_{\perp}}
 {\hat a}^{(\rw)\dagger}_{\Omega',\k_{\perp}'} |0_{M} \rangle
 = \f{1}{\l(1-\e^{-2\pi\Omega/a}\r)} \delta(\Omega-\Omega') \delta(\k_{\perp}-\k_{\perp}'),
\end{align}
and
\begin{align}
 \langle 0_{M}| {\hat a}^{(\rw)\dagger}_{\Omega,\k_{\perp}}
 {\hat a}^{(\rw)}_{\Omega',\k_{\perp}'} |0_{M} \rangle
 = \f{\e^{-2\pi\Omega/a}}{\l(1-\e^{-2\pi\Omega/a}\r)} \delta(\Omega-\Omega')
 \delta(\k_{\perp}-\k_{\perp}'),
\end{align}
we obtain
\begin{align}\label{eqn:AppendixB-CF}
 &\langle \delta {\hat X}_{\ud}(\eta) \delta {\hat X}_{\ud}(\eta') \rangle_{\rho^{(0)}}
 = -\f{\lambda^2}{8\pi^4a} \int_{\eta_{0}}^{\eta} \dd{{\tilde \eta}} \int_{\eta_{0}}^{\eta'} \dd{{\tilde \eta}'}
 \int_{-\infty}^{\infty} \d \Omega \int \d^2 \k_{\perp} K^2_{i\Omega/a}(k_{\perp}/a)
 \e^{\pi\Omega/a} \e^{-i\Omega ({\tilde \eta}-{\tilde \eta}')} \nn \\
 &\times \Tr_{\rho^{(0)}_{\cal E} \otimes \rho^{(0)}_{\cm}} \biggl\{\l[{\hat \mu}^{(0)}({\tilde \eta})
 \e^{ik_{x}{\hat x}^{(0)}({\tilde \eta})} \e^{ik_{y}{\hat y}^{(0)}({\tilde \eta})}, {\hat X}_{\ud}^{(0)}(\eta)\r]
 \times \l[{\hat \mu}^{(0)}({\tilde \eta}') \e^{-ik_{x}{\hat x}^{(0)}({\tilde \eta}')}
 \e^{-ik_{y}{\hat y}^{(0)}({\tilde \eta}')}, {\hat X}_{\ud}^{(0)}(\eta')\r]\biggr\},
\end{align}
where $\rho^{(0)} = \rho_{\cal E} \otimes \rho_{\cm} \otimes \ket{0_{M}}\bra{0_{M}}$, with
$\rho_{\cal E}$ and $\rho_{\cm}$ denote, respectively, the quantum states
corresponding to the internal energy levels, and CoM of the detector; and $\ket{0_{M}}\bra{0_{M}}$
denote the field, ${\hat \phi}$, in its Minkowski vacuum state.

Proceeding further, making use of the evolution equations
\begin{equation}
 {\hat \x}_{\perp}^{(0)}(\eta) = \f{{\hat \p}_{\perp}^{(0)}(\eta_{0})}{m} (\eta-\eta_{0})
 + {\hat \x}_{\perp}^{(0)}(\eta_{0}), ~~~~
 {\hat \p}^{(0)}_{\perp}(\eta) = {\hat \p}^{(0)}_{\perp}(\eta_{0}),
\end{equation}
and
\begin{align}
 {\hat \mu}^{(0)}(\eta) = {\hat \sigma}_{-} \e^{-i\omega_{0}\eta} + {\hat \sigma}_{+} \e^{i\omega_{0}\eta},
\end{align}
where ${\hat \sigma}_{-}=\ket{g}\bra{e}$ and ${\hat \sigma}_{+}=\ket{e}\bra{g}$, we obtain
in the limit $\eta_{0} \to -\infty$
\begin{align}
 & \Tr_{\rho^{(0)}_{\cal E} \otimes \rho^{(0)}_{\cm}} \l(\comm{{\hat \mu}({\tilde \eta})
 \e^{ik_{x}{\hat x}^{(0)}({\tilde \eta})} \e^{ik_{y}{\hat y}^{(0)}({\tilde \eta})}}{{\hat X}_{\ud}^{(0)}(\eta)}
 \times \comm{{\hat \mu}({\tilde \eta}') \e^{-ik_{x}{\hat x}^{(0)}({\tilde \eta}')}
 \e^{-ik_{y}{\hat y}^{(0)}({\tilde \eta}')}}{{\hat X}_{\ud}^{(0)}(\eta')}\r) \nn \\
 &= - \Tr_{\rho^{(0)}_{\cal E} \otimes \rho^{(0)}_{\cm}}
 \biggl\{\exp\l(-i\f{({\bm k}+{\hat \p}_{\perp})^2}{2m} ({\tilde \eta}-{\tilde \eta}')\r) \nn \\
 &\times \biggl[\e^{-i{\hat H}^{(0)}_{\cm} ({\tilde \eta}-{\tilde \eta}')}
 \times \l(\sigma_{-} \sigma_{+} \e^{i\omega_{0} ({\tilde \eta}-{\tilde \eta}')}
 + \sigma_{+} \sigma_{-} \e^{-i\omega_{0} ({\tilde \eta}-{\tilde \eta}')}\r)
 \times {\hat X}_{\ud}^{(0)}(\eta) {\hat X}_{\ud}^{(0)}(\eta') \nn \\
 &+ \e^{i{\hat H}^{(0)}_{\cm} ({\tilde \eta}-{\tilde \eta}')}
 \times \l(\sigma_{-} \sigma_{+} \e^{-i\omega_{0} ({\tilde \eta}-{\tilde \eta}')}
 + \sigma_{+} \sigma_{-} \e^{i\omega_{0} ({\tilde \eta}-{\tilde \eta}')}\r)
 \times {\hat X}_{\ud}^{(0)}(\eta') {\hat X}_{\ud}^{(0)}(\eta)\biggr]\biggr\}.
\end{align}

Making use of this in Eq.(\ref{eqn:AppendixB-CF}), and working in the regime where
$({\bm k}+\p_{\perp})^2/2m\ll a$, we obtain
\begin{align}
 &\langle \delta {\hat X}_{\ud}(\eta) \delta {\hat X}_{\ud}(\eta') \rangle_{\rho^{(0)}_{\cal E} \otimes \rho^{(0)}_{\cm}}
 = \f{\lambda^2}{8\pi^4a} \int_{0}^{\infty} \dd{{\tilde \eta}} \int_{0}^{\infty} \dd{{\tilde \eta}'}
 \int_{-\infty}^{\infty} \d \Omega \int \d^2 \k_{\perp} K^2_{i\Omega/a}(k_{\perp}/a)
 \e^{\pi\Omega/a} \e^{-i\Omega (\eta-\eta')} \e^{i\Omega ({\tilde \eta}-{\tilde \eta}')} \nn \\
 &\times \Tr_{\rho^{(0)}_{\cal E} \otimes \rho^{(0)}_{\cm}}
 \biggl\{\e^{-i{\hat H}^{(0)}_{\cm} (\eta-\eta')}
 \e^{i{\hat H}^{(0)}_{\cm} ({\tilde \eta}-{\tilde \eta}')}
 \biggl(\sigma_{-} \sigma_{+} \e^{i\omega_{0} (\eta-\eta')}
 \e^{-i\omega_{0} ({\tilde \eta}-{\tilde \eta}')}
 + \sigma_{+} \sigma_{-} \e^{-i\omega_{0} (\eta-\eta')}
 \e^{i\omega_{0} ({\tilde \eta}-{\tilde \eta}')}\biggr)
 \times {\hat X}^{(0)}(\eta) {\hat X}^{(0)}(\eta') \nn \\
 &+ \e^{i{\hat H}^{(0)}_{\cm} (\eta-\eta')} \e^{-i{\hat H}^{(0)}_{\cm} ({\tilde \eta}-{\tilde \eta}')}
 \biggl(\sigma_{-} \sigma_{+} \e^{-i\omega_{0} (\eta-\eta')}
 \e^{i\omega_{0} ({\tilde \eta}-{\tilde \eta}')}
 + \sigma_{+} \sigma_{-} \e^{i\omega_{0} (\eta-\eta')}
 \e^{-i\omega_{0} ({\tilde \eta}-{\tilde \eta}')}\biggr)
 \times {\hat X}^{(0)}(\eta') {\hat X}^{(0)}(\eta)\biggr\}.
\end{align}

Making the variable change, ${\tilde u}=a({\tilde \eta}-{\tilde \eta}')$
and ${\tilde v}=a({\tilde \eta}+{\tilde \eta}')/2$, and performing the $\k_{\perp}$ integral,
we calculate the correlation function, $S^{\ud}_{XX}(\eta-\eta')$, corresponding to the detector
observable, ${\hat X}_{\ud}$, as
\begin{align}
 & S^{\ud}_{XX}(\eta-\eta')
 \equiv \langle \delta {\hat X}_{\ud}(\eta) \delta {\hat X}_{\ud}(\eta') \rangle_{\rho^{(0)}_{\cal E}
 \otimes \rho^{(0)}_{\cm}} = \f{\lambda^2 T}{(2\pi)^3}
 \int_{-\infty}^{\infty} \dd {\tilde u} \Theta({\tilde u}) \int_{-\infty}^{\infty} \d \Omega
 \f{(\pi\Omega/a)}{\sinh(\pi\Omega/a)} \e^{\pi\Omega/a} \nn \\
 &\times \Tr_{\rho_{\ud}^{(0)}}
 \biggl\{\biggl(\sigma_{-} \sigma_{+} \e^{-i\l(\Omega-\omega_{0}+{\hat H}^{(0)}_{\cm}\r) (\eta-\eta')}
 \e^{i\l(\Omega-\omega_{0}+{\hat H}^{(0)}_{\cm}\r) {\tilde u}/a}
 + \sigma_{+} \sigma_{-} \e^{-i\l(\Omega+\omega_{0}+{\hat H}^{(0)}_{\cm}\r) (\eta-\eta')}
 \e^{i\l(\Omega+\omega_{0}+{\hat H}^{(0)}_{\cm}\r) {\tilde u}/a}\biggr) \nn \\
 &\times {\hat X}_{\ud}^{(0)}(\eta-\eta') {\hat X}_{\ud}^{(0)}(0) \nn \\
 &+ \biggl(\sigma_{-} \sigma_{+} \e^{-i\l(\Omega+\omega_{0}-{\hat H}^{(0)}_{\cm}\r) (\eta-\eta')}
 \e^{i\l(\Omega+\omega_{0}-{\hat H}^{(0)}_{\cm}\r) {\tilde u}/a}
 + \sigma_{+} \sigma_{-} \e^{-i\l(\Omega-\omega_{0}-{\hat H}^{(0)}_{\cm}\r) (\eta-\eta')}
 \e^{i\l(\Omega-\omega_{0}-{\hat H}^{(0)}_{\cm}\r) {\tilde u}/a}\biggr) \nn \\
 &\times {\hat X}_{\ud}^{(0)}(\eta'-\eta) {\hat X}_{\ud}^{(0)}(0)\biggr\}.
\end{align}
\section{Calculation details of Eq.(\ref{eqn:LabCF})}\label{Appendix:SYY}
The total Hamiltonian corresponding our detector-field setup in the laboratory frame can be written as
\begin{align}
 {\hat H} = \underbrace{\f{{\hat p}_{x}^2}{2m} + \f{{\hat p}_{y}^2}{2m}
 + \f{\omega_{0}}{2} \l(\dyad{e}{e} - \dyad{g}{g}\r)}_{{\hat H}_{S}}
 + \underbrace{\int \d^3 \k~\omega_{k} {\hat a}^{\dagger}_{\k} {\hat a}_{\k}}_{{\hat H}_{R}}
 - \underbrace{\lambda {\hat \mu}(t) {\hat \phi}(\tau,{\hat x},\hat{y},z)}_{\hat{H}_{\rm int}},
\end{align}
from which, using DDC formalism \cite{DDC1984} as stated in appendix.\ref{Appendix:DDC_Formalism},
we find the operator, ${\hat Y}_{\ud}$,
corresponding to the detector in the linear response regime as
\begin{align}
 \delta {\hat Y}_{\ud}(\tau) &= -\frac{i\lambda}{(2\pi)^{3/2}} \int_{\tau_{0}}^{\tau} \dd{{\tilde \tau}}
 \int \f{\d^3 \k}{\sqrt{2\omega_{k}}} \e^{ik_{z} z({\tilde \tau})}
 \biggl({\hat a}^{(0)}_{\k}(\tau_{0}) \e^{-i\omega_{k} t({\tilde \tau})}
 + {\hat a}^{(0)}_{-\k}(\tau_{0}) \e^{i\omega_{k} t({\tilde \tau})}\biggr) \nn \\
 &\times \l[{\hat \mu}^{(0)}({\tilde \tau}) \e^{ik_{x} {\hat x}^{(0)}({\tilde \tau})}
 \e^{ik_{y} {\hat y}^{(0)}({\tilde \tau})}, {\hat Y}_{\ud}^{(0)}(\tau)\r].
\end{align}

Making use of the expressions,
$\langle 0_{M}| {\hat a}_{\k} {\hat a}^{\dagger}_{\k} |0_{M} \rangle = \delta(\k-\k')$ and
$\langle 0_{M}| {\hat a}^{\dagger}_{\k} {\hat a}_{\k} |0_{M} \rangle = 0$, we obtain
\begin{align}
 \langle \delta {\hat Y}_{\ud}(\tau) \delta {\hat Y}_{\ud}(\tau') \rangle_{\rho^{(0)}}
 &= -\f{\lambda^2}{(2\pi)^{3}} \int_{\tau_{0}}^{t} \dd {\tilde \tau} \int_{\tau_{0}}^{\tau'} \dd {\tilde \tau}'
 \int \f{\d^3 \k}{2\omega_{k}} \e^{-i\omega_{k} {\tilde \tau}}
 \e^{i\omega_{k} {\tilde \tau}'} \e^{ik_{z}z({\tilde \tau})} \e^{-ik_{z}z({\tilde \tau}')} \nn \\
 &\times \Tr_{\rho^{(0)}_{\ud}} \biggl\{\comm{{\hat \mu}({\tilde \tau})
 \e^{ik_{x} {\hat x}^{(0)}({\tilde \tau})} \e^{ik_{y} {\hat y}^{(0)}({\tilde \tau})}}{{\hat Y}_{\ud}^{(0)}(\tau)}
 \comm{{\hat \mu}({\tilde \tau}') \e^{-ik_{x} {\hat x}^{(0)}({\tilde \tau}')}
 \e^{-ik_{y} {\hat y}^{(0)}({\tilde \tau}')}}{{\hat Y}_{\ud}^{(0)}(\tau')}\biggr\},
\end{align}
where $\rho^{(0)}=\rho^{(0)}_{\ud} \otimes \dyad{0_{M}}{0_{M}}$.
Proceeding further, we make use of the evolution equations
\begin{equation}
 {\hat \x}_{\perp}^{(0)}(\tau) = \f{{\hat \p}_{\perp}^{(0)}(\tau_{0})}{m} (\tau-\tau_{0})
 + {\hat \x}_{\perp}^{(0)}(\tau_{0}), ~~~~
 {\hat \p}^{(0)}_{\perp}(\tau) = {\hat \p}^{(0)}_{\perp}(\tau_{0}),
\end{equation}
and
\begin{align}
 {\hat \mu}^{(0)}(\tau) = {\hat \sigma}_{-} \e^{-i\omega_{0}\tau}
 + {\hat \sigma}_{+} \e^{i\omega_{0}\tau},
\end{align}
where ${\hat \sigma}_{-}=\ket{g}\bra{e}$ and ${\hat \sigma}_{+}=\ket{e}\bra{g}$, to obtain
\begin{align}
 \langle \delta {\hat Y}_{\ud}(\tau) \delta {\hat Y}_{\ud}(\tau') \rangle_{\rho^{(0)}}
 &= \f{\lambda^2}{(2\pi)^{3}} \int_{\tau_{0}=-\infty}^{t} \dd {\tilde \tau}
 \int_{\tau_{0}=-\infty}^{\tau'} \dd {\tilde \tau}'
 \int \f{\d^3 \k}{2\omega_{k}} \e^{-i\omega_{k} [t({\tilde \tau})-t({\tilde \tau}')]}
 \e^{i\f{k_{z}}{a} \l[z({\tilde \tau})-z({\tilde \tau})\r]} \nn \\
 &\times \Tr_{\rho_{\cm}^{(0)}}
 \biggl\{\e^{-i{\hat H}^{(0)}_{\cm} ({\tilde \tau}-{\tilde \tau}')}
 \times \l({\hat \sigma}_{-} {\hat \sigma}_{+} \e^{i\omega_{0} ({\tilde \tau}-{\tilde \tau}')}
 + {\hat \sigma}_{+} {\hat \sigma}_{-} \e^{-i\omega_{0} ({\tilde \tau}-{\tilde \tau}')}\r)
 \times {\hat Y}_{\ud}^{(0)}(\tau) {\hat Y}_{\ud}^{(0)}(\tau') \nn \\
 &+ \e^{i{\hat H}^{(0)}_{\cm} ({\tilde \tau}-{\tilde \tau}')}
 \times \l({\hat \sigma}_{-} {\hat \sigma}_{+} \e^{-i\omega_{0} ({\tilde \tau}-{\tilde \tau}')}
 + {\hat \sigma}_{+} {\hat \sigma}_{-} \e^{i\omega_{0} ({\tilde \tau}-{\tilde \tau}')}\r)
 \times {\hat Y}_{\ud}^{(0)}(\tau') {\hat Y}_{\ud}^{(0)}(\tau)\biggr\},
\end{align}
in the limit $\tau_{0} \to -\infty$ and $({\bm k}+\p_{\perp})^2/2m\ll a$.

Making use of the trajectory information of the detector, $t=a^{-1} \sinh a\tau$,
$z=a^{-1} \sinh a\tau$, we obtain
\begin{align}
 & S^{\ud}_{YY}(\tau-\tau') \equiv \langle \delta {\hat Y}_{\ud}(\tau)
 \delta {\hat Y}_{\ud}(\tau') \rangle_{\rho^{(0)}}
 = \f{\lambda^2}{(2\pi)^{3}} \int_{0}^{\infty} \dd {\tilde \tau}
 \int_{0}^{\infty} \dd {\tilde \tau}' \int \d^2 \k_{\perp}
 \int_{-\infty}^{\infty} \f{\d k_{z}}{2\omega_{k}}
 \e^{-i\f{\omega_{k}}{2a} \sinh [a(\tau-\tau'-{\tilde \tau}+{\tilde \tau}')/2]} \nn \\
 &\times \Tr_{\rho_{\cm}^{(0)}}
 \biggl\{\e^{-i{\hat H}^{(0)}_{\cm} (\tau-\tau')}
 \e^{i{\hat H}^{(0)}_{\cm} ({\tilde \tau}-{\tilde \tau}')}
 \times \biggl({\hat \sigma}_{-} {\hat \sigma}_{+} \e^{i\omega_{0} (\tau-\tau')}
 \e^{-i\omega_{0} ({\tilde \tau}-{\tilde \tau}')}
 + {\hat \sigma}_{+} {\hat \sigma}_{-} \e^{-i\omega_{0} (\tau-\tau')}
 \e^{i\omega_{0} ({\tilde \tau}-{\tilde \tau}')}\biggr)
 {\hat Y}_{\ud}^{(0)}(\tau) {\hat Y}_{\ud}^{(0)}(\tau') \nn \\
 &+ \e^{i{\hat H}^{(0)}_{\cm} (\tau-\tau')}
 \e^{-i{\hat H}^{(0)}_{\cm} ({\tilde \tau}-{\tilde \tau}')}
 \times \biggl({\hat \sigma}_{-} {\hat \sigma}_{+} \e^{-i\omega_{0} (\tau-\tau')}
 \e^{i\omega_{0} ({\tilde \tau}-{\tilde \tau}')}
 + {\hat \sigma}_{+} {\hat \sigma}_{-} \e^{i\omega_{0} (\tau-\tau')}
 \e^{-i\omega_{0} ({\tilde \tau}-{\tilde \tau}')}\biggr)
 {\hat Y}_{\ud}^{(0)}(\tau') {\hat Y}_{\ud}^{(0)}(\tau)\biggr\},
\end{align}
and performing a variable change ${\tilde u}={\tilde \tau}-{\tilde \tau}'$ and
${\tilde v}=({\tilde \tau}+{\tilde \tau}')/2$ , we obtain the correlation function,
$S^{\ud}_{YY}(\tau-\tau')$, as
\begin{align}
 & S^{\ud}_{YY}(\tau-\tau') = \f{\lambda^2 T}{(2\pi)^{3}} \int_{0}^{\infty} \dd {\tilde u}
 \int \d^2 \k_{\perp} \int_{-\infty}^{\infty} \f{\d k'_{z}}{2\omega'_{k}}
 \e^{-i\f{\omega'_{k}}{2a} \sinh [a(\tau-\tau'-{\tilde u})/2]} \nn \\
 &\times \Tr_{\rho_{\cm}^{(0)}}
 \biggl\{\e^{-i{\hat H}^{(0)}_{\cm} (\tau-\tau')} \e^{i{\hat H}^{(0)}_{\cm} {\tilde u}}
 \biggl({\hat \sigma}_{-} {\hat \sigma}_{+} \e^{i\omega_{0} (\tau-\tau')} \e^{-i\omega_{0} {\tilde u}}
 + {\hat \sigma}_{+} {\hat \sigma}_{-} \e^{-i\omega_{0} (\tau-\tau')} \e^{i\omega_{0} {\tilde u}}\biggr)
 {\hat Y}_{\ud}^{(0)}(\tau) {\hat Y}_{\ud}^{(0)}(\tau') \nn \\
 &+ \e^{i{\hat H}^{(0)}_{\cm} (\tau-\tau')}
 \e^{-i{\hat H}^{(0)}_{\cm} {\tilde u}}
 \biggl({\hat \sigma}_{-} {\hat \sigma}_{+} \e^{-i\omega_{0} (\tau-\tau')}
 \e^{i\omega_{0} {\tilde u}}
 + {\hat \sigma}_{+} {\hat \sigma}_{-} \e^{i\omega_{0} (\tau-\tau')}
 \e^{-i\omega_{0} {\tilde u}}\biggr)
 {\hat Y}_{\ud}^{(0)}(\tau') {\hat Y}_{\ud}^{(0)}(\tau)\biggr\}.
\end{align}

Evaluating the $k_{z}$ integral, and rewriting appropriately, we obtain
\begin{align}
 & S^{\ud}_{YY}(\tau-\tau') = \f{\lambda^2}{(2\pi)^{3}} \int \d^2 \k_{\perp} \int_{-\infty}^{\infty}
 \dd {\tilde u} \Theta({\tilde u}) \int_{-\infty}^{\infty} \dd \Omega'
 \l(\f{2}{a}\r) \e^{\pi(\Omega'/a)} K^2_{i\Omega'/a}(k_{\perp}/a) \nn \\
 &\times \Tr_{\rho_{\cm}^{(0)}}
 \biggl\{\biggl[{\hat \sigma}_{-} {\hat \sigma}_{+} \e^{-i(\Omega'-\omega_{0}+{\hat H}^{(0)}_{\cm}) (\tau-\tau')}
 \e^{i(\Omega'-\omega_{0}+{\hat H}^{(0)}_{\cm}) {\tilde u}}
 + {\hat \sigma}_{+} {\hat \sigma}_{-} \e^{-i(\Omega'+\omega_{0}+{\hat H}^{(0)}_{\cm}) (\tau-\tau')}
 \e^{i(\Omega'+\omega_{0}+{\hat H}^{(0)}_{\cm}) {\tilde u}}\biggr]
 {\hat Y}_{\ud}^{(0)}(\tau-\tau') {\hat Y}_{\ud}^{(0)}(0) \nn \\
 &+ \biggl[{\hat \sigma}_{-} {\hat \sigma}_{+} \e^{-i(\Omega'+\omega_{0}-{\hat H}^{(0)}_{\cm}) (\tau-\tau')}
 \e^{i(\Omega'+\omega_{0}-{\hat H}^{(0)}_{\cm}) {\tilde u}}
 + {\hat \sigma}_{+} {\hat \sigma}_{-} \e^{-i(\Omega'-\omega_{0}-{\hat H}^{(0)}_{\cm}) (\tau-\tau')}
 \e^{i(\Omega'-\omega_{0}-{\hat H}^{(0)}_{\cm}) {\tilde u}}\biggr]
 {\hat Y}_{\ud}^{(0)}(\tau'-\tau) {\hat Y}_{\ud}^{(0)}(0) \biggr\},
\end{align}
and further evaluating the $k_{\perp}$ integral leads to
\begin{align}
 & S_{XX}(u) = \f{\lambda^2}{(2\pi)^{2}} \int_{-\infty}^{\infty}
 \dd {\tilde u} \Theta({\tilde u}) \int_{-\infty}^{\infty} \dd \Omega'
 \e^{\pi(\Omega'/a)} \f{(\pi\Omega'/a)}{\sinh(\pi\Omega'/a)} \nn \\
 &\times \Tr_{\rho_{\cm}^{(0)}}
 \biggl\{\biggl[{\hat \sigma}_{-} {\hat \sigma}_{+} \e^{-i(\Omega'-\omega_{0}+{\hat H}^{(0)}_{\cm}) (\tau-\tau')}
 \e^{i(\Omega'-\omega_{0}+{\hat H}^{(0)}_{\cm}) {\tilde u}/a}
 + {\hat \sigma}_{+} {\hat \sigma}_{-} \e^{-i(\Omega'+\omega_{0}+{\hat H}^{(0)}_{\cm}) (\tau-\tau')}
 \e^{i(\Omega'+\omega_{0}+{\hat H}^{(0)}_{\cm}) {\tilde u}/a}\biggr]
 {\hat Y}_{\ud}^{(0)}(\tau-\tau') {\hat Y}_{\ud}^{(0)}(0) \nn \\
 &+ \biggl[{\hat \sigma}_{-} {\hat \sigma}_{+} \e^{-i(\Omega'+\omega_{0}-{\hat H}^{(0)}_{\cm}) (\tau-\tau')}
 \e^{i(\Omega'+\omega_{0}-{\hat H}^{(0)}_{\cm}) {\tilde u}/a}
 + {\hat \sigma}_{+} {\hat \sigma}_{-} \e^{-i(\Omega'-\omega_{0}-{\hat H}^{(0)}_{\cm}) (\tau-\tau')}
 \e^{i(\Omega'-\omega_{0}-{\hat H}^{(0)}_{\cm}) {\tilde u}/a}\biggr]
 {\hat Y}_{\ud}^{(0)}(\tau'-\tau) {\hat Y}_{\ud}^{(0)}(0)\biggr\}.
\end{align}
\end{widetext}

\end{document}